# Propensity score matching for multiple treatment levels: A CODA-based contribution

**Hajime Seya**[*1]

Graduate School of Engineering Faculty of Engineering, Kobe University,

1-1 Rokkodai-cho, Nada-ku, Kobe 657-8501, Japan,

E-mail: hseya@people.kobe-u.ac.jp

**Takahiro Yoshida**

Graduate School of Systems and Information Engineering, University of Tsukuba,

1-1-1, Tennodai, Tsukuba, Ibaraki, 305-8573, Japan,

E-mail: yoshida.takahiro@sk.tsukuba.ac.jp

**Abstract**

This study proposes a simple technique for propensity score matching for multiple treatment levels under the strong unconfoundedness assumption with the help of the Aitchison distance proposed in the field of compositional data analysis (CODA).

Keywords: propensity score matching, multiple treatment group, compositional data analysis (CODA), Aitchison distance, strong unconfoundness

JEL: C4; C5

Acknowledgement: This study was funded by JSPS KAKENHI Grant Number 17K14738 and 15H04054.

## 1. Introduction

The propensity score matching (PSM), propensity score weighting (PSW), and propensity score subclassification (PSS) play important roles in causal inference across disciplines, including economics. However, most of the works focus on binary setting, paying less attention on the setting with more than two treatment levels. Moreover, most of the existing studies on multiple treatment levels focus on the PSW or PSS (Imai and van Dyk, 2004; McCaffrey et al., 2013) rather than the PSM, except for a few seminal attempts (Imbens, 2000; Yang et al., 2016). The present study proposes a simple PSM approach for multiple treatment levels with the help of the Aitchison distance proposed in the field of compositional data analysis (CODA).

---

[1] *: (Corresponding author)

## 2. Method

Our framework follows Imbens (2000) and Yang et al. (2016). The treatment for unit *i* is denoted by $T_i \in \{1, ..., T\}$, with $T > 2$. For the unit *i*, we define potential outcome $Y(T_i)$, and observed outcome $Y_i^{obs} = Y(T_i = t)$, where the latter corresponds to the treatment received. In addition, we define a vector of pre-treatment variables denoted by $X_i$. With these settings, we introduce Imbens' (2000) generalized propensity score as follows:

**DEFINITION 1** *Generalized propensity score. The generalized propensity score is the conditional probability of receiving a particular level of treatment given the pre-treatment variables.*
$$p(t|x) = pr(T_i = t|X_i = x).$$
The generalized propensity score is assumed to be $0 < p(t|x) < 1$, for all *t*, *x*, and $\sum_{t=1}^{T} p(t|x) = 1$.

Next, following Rosenbaum and Rubin (1983) and Yang et al. (2016), we introduce the concept of strong unconfoundedness.

**DEFINITION 2** *Strong unconfoundedness.* [2] *The assignment mechanism is strongly unconfounded if*
$$T_i \perp (Y_i(1), ..., Y_i(T))|X_i.$$
Here, $\perp$ denotes independence. Definition 2 implies the following lemma 1 (Rosenbaum and Rubin, 1983; Yang et al., 2016).

**LEMMA 1**
$$T_i \perp (Y_i(1), ..., Y_i(T))|(p(1|X_i), ..., p(T-1|X_i))'.$$
Here, $p(T|X_i)$ is omitted from conditioning set because $p(T|X_i)$ can be obtained as a linear combination of $p(1|X_i), ..., p(T-1|X_i)$.

In the binary treatment case (let 1: treated; 0: control (untreated)), lemma 1 falls to $T_i \perp (Y_i(1), Y_i(0))|p(1|X_i)$, it follows that independence between $T_i$ and $(Y_i(1), Y_i(0))$ can be satisfied by not directly balancing (possibly high dimensional) $X_i$, but indirectly balancing $X_i$ by projecting $X_i$ to a scalar "balancing score" $b(X_i)$ and balance $X_i$ through $b(X_i)$. In case of the PSM, the balancing score is given as the propensity score $b(X_i) = p(1|X_i)$. Then, if treatment unit and a corresponding matched unit in the control group have the same propensity score, the two matched units will have, supposedly, the same value of covariate vector $X_i$. Typically, the estimate of $p(1|X_i)$ is obtained by applying the logit/probit model. The average treatment effect on the treated (ATT), which is our interest, can be obtained as
$$\tau_{ATT} = E\{E[Y(1)|P(1|X_i)] - E[Y(0)|P(0|X_i)]\} = E\{E[Y(1) - Y(0)|P(1|X_i)]\}. \quad (1)$$
Thus, ATT can be obtained simply as the mean difference in outcomes, appropriately weighted by the propensity score distribution (Caliendo and Kopeinig, 2008).

---
[2] Imbens (2000) proposes weaker condition of unconfoundedness. See Yang et al. (2016) regarding the difference between strong and weak unconfoundedness. They propose PSM approach under the weak unconfoundedness assumption.

However, in multiple treatment groups setting, independence between $T_i$ and $(Y_i(1), \ldots, Y_i(T))$ can be satisfied conditioning on a propensity score vector $(p(1|X_i), \ldots, p(T-1|X_i))'$, but not on a scalar score. It follows that the ATT, under strongly uncoufoundedness assumption, may be given as

$$\tau_{ATT} = E\{E[Y(t) - Y(s) | (p(1|X_i), \ldots, p(T-1|X_i))']\}; \quad t, s \in T. \tag{2}$$

Yang et al. (2016) and references therein suggest that without additional assumptions, there is generally no scalar balancing score $b(x)$ such that $T_i \perp (Y_i(1), \ldots, Y_i(T)) | b(X_i)$. Hence, instead of finding $b(x)$ with additional assumptions, we follow a different approach, as summarized in the next section.

## 3. CODA-based PSM approach

Let the propensity score vector for unit $i$ given as $\boldsymbol{p}_i \equiv (p(1|X_i), \ldots, p(T|X_i))'$. Also, let $Y_i^{obs} = Y(T_i = t)$ and $\boldsymbol{Y}_{\boldsymbol{j}}^{obs} = Y(T_i = s)$, where the subscript in bold $\boldsymbol{j} \equiv (j_1, \ldots, j_M)'$ ($M = \|\boldsymbol{j}\|$) denotes the set of unit on which the treatment $s$ is conducted for $Y$. The estimate for $\boldsymbol{p}_i$, say $\hat{\boldsymbol{p}}_i$, can be obtained by applying the multinomial probit model or other methods. Although there are many propensity-based matching algorithms (Caliendo and Kopeinig, 2008; Huber et al., 2013), they typically require to measure the "distance" between $\hat{\boldsymbol{p}}_i$ and $\hat{\boldsymbol{p}}_j$, and find the optimal subset of the unit $\boldsymbol{j}$ based on some objective function.

Incidentally, any vector $\boldsymbol{z} = (z_1, \ldots, z_D)'$ representing proportions of some whole that is subject to the sum constraint $(z_1 + z_2, \ldots + z_D)$ is called compositional data. Clearly, $\hat{\boldsymbol{p}}_i$ and $\hat{\boldsymbol{p}}_{j_m}$ are compositional data, because by definition of generalized propensity score, $\sum_{t=1}^{T} \hat{p}_i(t|x) = 1$ and $\sum_{t=1}^{T} \hat{p}_{j_m}(t|x) = 1$ hold. The analysis for compositional data, say CODA, had initially been developed especially in geology (see Aitchison, 1986), but gradually it began to be used in various fields. Findings in CODA literature suggest that when we measure the difference between compositional data, the Euclidean distance $d^{Euc}(\hat{\boldsymbol{p}}_i, \hat{\boldsymbol{p}}_{j_m}) = \sqrt{\sum_{t=1}^{T} (\hat{p}_i(t) - \hat{p}_{j_m}(t))^2}$ may be irrelevant (Aitchison et al., 2000). Let us take a simple example illustrated in Pawlowsky-Glahn et al. (2015). Consider the compositions [5, 65, 30], [10, 60, 30], [50, 20, 30], and [55, 15, 30]. Intuitively, the difference between [5, 65, 30] and [10, 60, 30] is not the same as the difference between [50, 20, 30] and [55, 15, 30]. However, the Euclidean distance between them is certainly the same. In the first case, the proportion in the first component is doubled, while in the second case, the relative increase is about 10%. This "relative difference" seems more adequate to describe compositional variability. The Aitchison distance (Aitchison, 1986), defined as $d^{Ait}(\hat{\boldsymbol{p}}_i, \hat{\boldsymbol{p}}_{j_m}) = d^{Euc}(clr(\hat{\boldsymbol{p}}_i), clr(\hat{\boldsymbol{p}}_{j_m}))$ (where $clr(\boldsymbol{z}) = [\ln\frac{z_1}{g(\boldsymbol{z})}, \ldots, \ln\frac{z_D}{g(\boldsymbol{z})}]$; with $g(\boldsymbol{z}) = (\prod_{d=1}^{D} z_d)^{1/D}$) satisfy this requirement.[3] That is, it produces the same distance between 0.01 and 0.02 and between 0.1 and

---

[3] The Aitchison distance is ideal as the distance measure for compositional data, in a sense, satisfies not only usual conditions as a distance measure (i.e., non-negativity, symmetry, and triangle condition), but also the four additional conditions proposed by Aitchison (1992) (scale invariance, permutation invariance, perturbation invariance, and subcompositional coherence). See Aitchison (1992) for details.

0.2.

Now, imagine very high dimensional compositions. Naturally, the share for each component will be small. In such a case, the Euclidean distance between any two such compositions will be very small as well, even though they may have components that are many-fold different in relative abundance. The Aitchison distance, with its focus on the ratio of corresponding components, will emphasize these differences in relative abundance much more effectively (Lovell et al., 2011). However, the choice of distance measure is left to researchers (Otero et al., 2005). If we judge that "relative difference" is not adequate, we can use the Euclidean distance.

After matching, ATT can be obtained simply as the mean difference in outcomes, appropriately weighted by the propensity score distribution, just as in the binary case. The above procedure is very simple and easy to implement with standard software such as R[4] or Stata.

## 4. Conclusions

We proposed PSM for multiple treatment levels under the strong unconfoundedness assumption (Rosenbaum and Rubin, 1983; Yang et al., 2016). Matching may be conducted with the Euclidean distance or the Aitchison distance, where the latter is formalized for compositional data.

---

[4] For instance, aDist function of robCompositions package in R can be available for computing the Aitchison distance.